\documentclass[twocolumn]{jpsj2} 
\title{Reexamination of Finite-Lattice Extrapolation of Haldane Gaps}
\catcode`\@=11
\def\simle{\mathrel{\mathpalette\@versim<}}   
\def\simge{\mathrel{\mathpalette\@versim>}}   
\def\@versim#1#2{\lower2.5pt\vbox{\baselineskip0pt \lineskip-.5pt
   \ialign{$\m@th#1\hfil##\hfil$\crcr#2\crcr\sim\crcr}}}
\catcode`\@=12

\author{
Hiroki \textsc{Nakano}
\thanks{E-mail address: hnakano@sci.u-hyogo.ac.jp} 
and 
Akira \textsc{Terai}${}^{1}$
\thanks{E-mail address: terai@a-phys.eng.osaka-cu.ac.jp}
}

\inst{%
Graduate School of Material Science, University of Hyogo,
Kouto 3-2-1, Kamiori, Ako-gun, 678-1297 \\
${}^{1}$Department of Applied Physics, Osaka City University, 
Sumiyoshi-ku, Osaka 558-8585
}

\recdate{\today}

\abst{We propose two methods of estimating a systematic error 
in extrapolation to the infinite-size limit 
in the study of measuring the Haldane gaps 
of the one-dimensional Heisenberg antiferromagnet 
with the integer spin up to $S=5$. 
The finite-size gaps obtained by numerical diagonalizations 
based on Lanczos algorithm are presented 
for sizes that have not previously been reported. 
The changes of boundary conditions are also examined. 
We successfully demonstrate that our methods of extrapolation 
work well. 
The Haldane gap for $S=1$ is estimated to be 
$0.4104789 \pm 0.0000013$.  
We successfully obtain the gaps up to $S=5$, 
which make us confirm the asymptotic formula of the Haldane gap 
in $S\rightarrow\infty$. 
}

\kword{%
Antiferromagnetic Heisenberg spin chain,
Haldane gap, 
Exact-diagonalization method, 
Lanczos method, 
Extrapolation
}

\begin{document}
\maketitle

\section{Introduction} 

Extrapolation is a fundamental technique 
in a lot of studies in physics. 
In particular, the technique is often carried out 
in the condensed-matter physics when one attempts to know 
a quantity in the thermodynamic limit 
from several finite-size data. 
One of the reliable ways to obtain such finite-size data 
is the numerical-diagonalization method applied to 
the Hamiltonian matrix describing a system. 
This method provides us with very precise finite-size data 
although available system sizes are limited 
to being very small. 
This method is non-biased against the effects of interaction; 
thus it contributes much to the understanding 
of many-body problems. 
Typical examples are the quantum spin systems, 
in which there often appear nontrivial quantum states 
due to the presence of interactions between spins. 
One of them is the ground state 
of the integer-spin one-dimensional Heisenberg antiferromagnet. 
In this system, an energy gap exists 
between the unique ground state and the first excited state; 
this gap is called the Haldane gap\cite{Haldane1,Haldane2}. 

The magnitude of the Haldane gap has been estimated 
by various numerical methods\cite{history_Haldane_gap}. 
There are three representative approaches.  
The first one is the numerical Lanczos diagonalization 
of finite-size clusters. 
In the $S=1$ case, system sizes up to 22 sites were treated 
under the periodic boundary condition\cite{Golinelli2}. 
Since the available system sizes are small, 
an appropriate extrapolation is required. 
Unfortunately, it is difficult to estimate 
a systematic error due to the extrapolation. 
The second one is the density matrix renormalization group 
(DMRG) method\cite{DMRG}. 
The calculation was carried out 
under a peculiar boundary condition, namely, 
each edge of the $S=1$ chain connecting 
with an $S=1/2$ spin. 
In this way, it is necessary to tune 
the artificial interaction at the edges. 
The third one is a quantum Monte Carlo (QMC) 
simulation\cite{Todo_Kato_QMC}. 
Since this simulation was performed by the loop algorithm 
together with a continuous imaginary time technique, 
calculations of very large systems are available. 
However, a statistical error due to a Monte Carlo sampling 
cannot be avoided. 
Up to the present time, these approaches give 
consistent estimates of the Haldane gap 
with their own errors of the same order. 

Under such circumstances, we attempt again 
to estimate the Haldane gaps 
of the Heisenberg antiferromagnetic spin chain 
as precisely as possible. 
Our main method is the numerical diagonalization. 
A primary purpose of this paper is to propose a procedure 
to obtain a reliable error in extrapolation 
of finite-size data toward the infinite size of the system. 

In the extrapolation, the weak system-size dependence 
of the finite-size gap is favored. 
The system-size dependence of the gap is determined 
by the choice of the boundary condition. 
It is known that 
the dependence becomes suppressed 
when one twists the boundary condition 
from the periodic boundary condition\cite{Terai}.  
Under this background, 
we examine boundary conditions 
in the case of $S=1$ 
to know which boundary condition is appropriate. 
We then find that the twisted boundary condition is 
the most appropriate one 
among periodic, twisted, and open boundary conditions. 
The twisted boundary condition gives 
a good sequence of finite-size excitation gaps 
for the sake of the extrapolation. 
Next, we develop a method to obtain a reliable error 
when the convergence of the data sequence can been accelerated. 
Thereby, a very precise estimation of the $S=1$ Haldane gap 
is successfully obtained. 
When one imposes the twisted boundary condition 
for $S=2$, 3, 4, and 5, the Haldane gaps can be obtained 
to be nonzero values in the thermodynamic limit 
in spite of the fact that the Lanczos method can treat 
only the extremely small system sizes. 
When $S$ becomes larger, 
the convergence acceleration becomes more difficult. 
In such a case of the acceleration in failure, 
we also develop another procedure to estimate an error 
to the excitation gap of the infinite system. 

This paper is organized as follows. 
In the next section, the model Hamiltonian and the method of 
calculation will be explained. 
In the first half of \S 3, 
the numerical results of the $S=1$ system are presented 
and discussed. 
Boundary conditions are examined and 
the extrapolation based on the convergence acceleration 
is performed. 
We propose its error estimation and demonstrate the validity. 
In the second half of \S 3, 
the cases of $S\ge 2$ are studied. 
Another procedure of obtaining an error is introduced. 
The final section is devoted to the summary and some remarks. 

\section{Hamiltonian and Method}

The Hamiltonian of the present model is given by 
\begin{equation}
{\cal H} = \sum_{i=1}^{N} J_{m} 
\mbox{\boldmath $S$}_{m}\cdot\mbox{\boldmath $S$}_{m+1},
\end{equation}
where $\mbox{\boldmath $S$}_{m}$ is a spin operator 
with its amplitude $S$ at site $m$. 
Here $N$ is the number of spin sites. 
The system size $N$ is supposed to be an even integer. 
The amplitude of the exchange interaction is denoted by $J_{m}$ 
which will be defined later 
when the boundary conditions is explained. 

In this paper, we carry out numerical diagonalizations 
for finite-size clusters of systems of $S=1$, 2, 3, 4, and 5. 
We calculate the ground-state energy $E_{0}$ and 
excitation energies (the first excitation $E_{1}$ and 
the second excitation $E_{2}$) 
by the method of Lanczos algorithm\cite{Householder}. 
We have successfully developed a code for parallel calculations 
of the Lanczos algorithm. 
The maximum sizes in this paper are 
$N=24$ for $S=1$, 
$N=16$ for $S=2$, 
$N=12$ for $S=3$, 
$N=12$ for $S=4$, and 
$N=10$ for $S=5$. 
These sizes have not been treated in the Lanczos calculations 
as long as the present authors know. 
Note that we only assume the conservation 
of the $z$-component of the total spin. 
Thus, arbitrary shapes of clusters can be treated. 
The dimensions of the Hilbert space are very large. 
For example, the dimension of the largest subspace, 
namely $S_{\rm tot}^{z}=0$, of $N=24$ for $S=1$ 
is $27~948~336~381$, 
where $S_{\rm tot}^{z}$ is the $z$ component of the total spin. 
It is worth to emphasize that 
the parallelization makes it possible to carry out 
the Lanczos calculations. 

\section{Result and Discussion}

\subsection{Case for $S=1$}

\subsubsection{Boundary Conditions}

Let us consider differences from the choice 
of boundary conditions.
The differences from boundary conditions appear in results 
of calculations of finite-size systems. 
On the other hand, the differences are supposed to disappear 
in the limit of $N\rightarrow\infty$. 
Finite-size effects depend on the choice of a boundary condition. 
Namely, a different type of boundary condition gives 
a different finite-size sequence concerning 
with a physical quantity. 
In order to obtain precisely the information 
in the thermodynamic limit 
that is not affected by boundary conditions, 
one should employ an appropriate boundary condition. 
Such a condition is not necessarily 
the periodic boundary condition. 
The appropriate condition depends 
on systems and physical quantities. 
Therefore, the examination of various boundary conditions is 
important. 
Here we focus our attention on the problem of the Haldane gaps 
and begin with such an examination in the $S=1$ case. 

In this paper, we examine three types of the boundary conditions: 
the open, periodic, and twisted boundary conditions.
The open boundary condition is given by 
\begin{equation}
J_{m}=
\left\{
\begin{array}{ll}
1 & (m<N) 
\\
0 & (m=N).
\end{array}
\right.
\end{equation}
The periodic boundary condition is given by 
$\mbox{\boldmath $S$}_{N+1}=\mbox{\boldmath $S$}_{1}$ 
and $J_{m}=1$ for arbitrary $m$. 
The twisted boundary condition 
is given by 
\begin{equation}
S^{x}_{N+1}=-S^{x}_{1}, \ 
S^{y}_{N+1}=-S^{y}_{1}, \ 
S^{z}_{N+1}= S^{z}_{1}, 
\end{equation}
and $J_{m}=1$ for arbitrary $m$. 
Note here that energies are measured in units of 
nonzero $J_{m}$; therefore we take it unity. 

\begin{figure}[tb]
\begin{center}
\includegraphics[width=8cm]{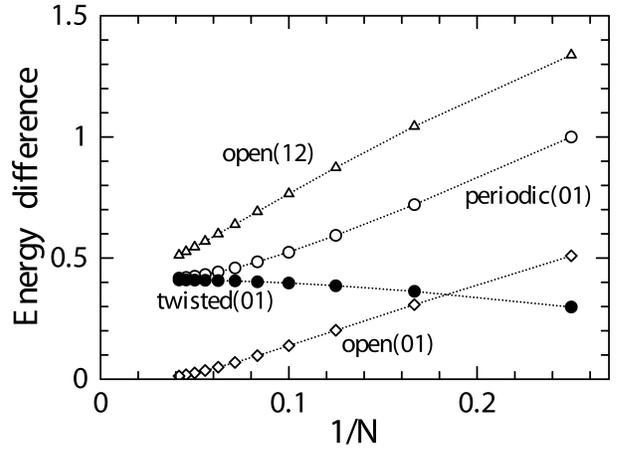}
\end{center}
\caption{Energy differences of finite-size $S=1$ systems. 
Open and closed circles denote results of $E_{1}-E_{0}$
under the periodic boundary condition and 
the twisted boundary condition, respectively. 
Triangles and diamonds denote results of $E_{2}-E_{1}$ and 
$E_{1}-E_{0}$ under the open boundary condition, respectively. 
Note that the maximum system size is $N=24$ 
irrespective of boundary conditions. 
}
\label{bc-compr}
\end{figure}

\begin{figure}[tb]
\begin{center}
\includegraphics[width=8cm]{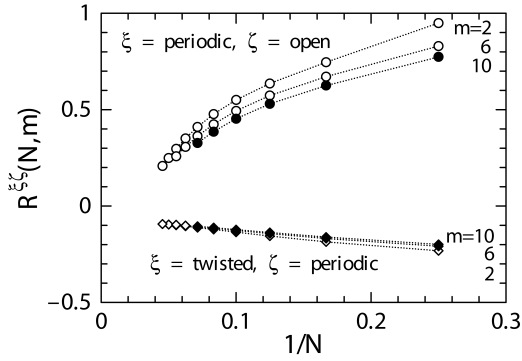}
\end{center}
\caption{Ratio $R^{\xi\zeta}(N,m)$ v.s. inversed system sizes. 
Circles denote the case 
of $\xi={\rm periodic}, \zeta={\rm open}$. 
Diamonds denote the case 
of $\xi={\rm twisted}, \zeta={\rm periodic}$. 
Closed symbols mean $m=10$ which is the largest $m$ 
in this figure.  
}
\label{ratio_fig}
\end{figure}

First, we show numerical results of system size dependence 
of energy differences in Fig.~\ref{bc-compr}. 
Our results under the periodic boundary condition 
agree with those in ref.~\ref{Golinelli2} up to $N=22$; 
our results for $N=24$ are new.    
One can clearly observe that $E_{1}-E_{0}$ 
under the open boundary condition vanishes 
in the limit of $N\rightarrow\infty$. 
This behavior is consistent with a quasi-degeneracy 
of the Haldane-type ground states. 
Under the open boundary condition, the Haldane gap appears 
above these degenerate ground states; 
the energy difference $E_{2}-E_{1}$ decreases gradually 
when $N$ is increased and seems to converge around 0.4. 
Under the twisted boundary condition, on the other hand, 
the energy difference $E_{1}-E_{0}$ increases 
with increasing $N$. 
This dependence will also be very useful 
when we study gaps for the cases of $S \ge 2$. 
In this section, let us compare the speed of the convergence 
of the finite-size sequence under each boundary condition. 
In order to achieve it, we consider the ratio defined as 
\begin{equation}
R^{\xi\zeta}(N,m) 
= 
\frac{G^{\xi  }(N)-G^{\xi  }(N+m)}{
      G^{\zeta}(N)-G^{\zeta}(N+m)}. 
\label{define_ratio}
\end{equation}
Here $G^{\zeta}(N)$ is the energy difference 
with system size $N$ under the $\zeta$ boundary condition 
which converges to the Haldane gap in the limit 
of $N\rightarrow\infty$; 
namely, $G^{\zeta}(N)=E_{1}-E_{0}$ for 
the periodic and twisted boundary conditions, 
and $G^{\zeta={\rm open}}(N)=E_{2}-E_{1}$. 
Results of the ratio $R^{\xi\zeta}(N,m)$ 
v.s. $1/N$ are depicted for two types in Fig.~\ref{ratio_fig}. 
All the absolute values of the presented ratios are 
less than unity. 
The ratio of $\xi={\rm periodic}$ and $\zeta={\rm open}$ 
gets gradually smaller when $N$ is increased. 
The ratio seems to vanish in the limit of $N\rightarrow\infty$. 
If the ratio in the limit of $m\rightarrow\infty$ vanishes 
and when each $G^{\zeta}(N)$ converges to the same value, 
namely the Haldane gap $\Delta(S=1)$ in this case, 
one obtains 
\begin{equation}
\lim_{N\rightarrow\infty}
\frac{G^{\xi  }(N)-\Delta(S=1)}{
      G^{\zeta}(N)-\Delta(S=1)}
=0.
\end{equation}
This suggests that 
the sequence under the periodic boundary condition converges 
faster than that under the open boundary condition. 
(See appendix.) 
Thus the periodic boundary condition is more appropriate 
than the open boundary condition to measure the Haldane gap 
within the system sizes that are available 
in numerical diagonalization calculations. 
On the other hand, 
the ratio of $\xi={\rm twisted}$ and $\zeta={\rm periodic}$ 
in the limit of $N\rightarrow\infty$ 
seems not to vanish but to converge to a nonzero value. 
This means that 
concerning with the convergence of the sequence, 
the speeds of cases of the twisted and periodic 
boundary conditions are almost the same with each other. 
Even though the speeds are comparable, 
the absolute values of the ratios are much smaller than unity 
($|R^{\xi\zeta}(N,m)|<1/4$). 
This suggests that each datum of the sequence 
under the twisted boundary condition is closer to 
the Haldane gap $\Delta(S=1)$ than the corresponding datum 
under the periodic boundary condition. 
In this meaning, we can conclude that 
the twisted boundary condition is more appropriate 
than the periodic boundary condition. 
Therefore, the twisted boundary condition is the most appropriate 
among the present three conditions. 

\subsubsection{Extrapolation}

In this subsection, we attempt to extrapolate the above sequences 
of the $S=1$ finite-size energy differences by means of the technique 
of convergence acceleration. 
We here apply Wynn's epsilon algorithm\cite{Wynn} given by
\begin{eqnarray}
\frac{1}{A^{(k+1)}_{N}-A^{(k)}_{N-2}}
&=&
\frac{1}{A^{(k)}_{N-4}-A^{(k)}_{N-2}}
+\frac{1}{A^{(k)}_{N}-A^{(k)}_{N-2}}
\nonumber \\
& &
-\frac{\alpha}{A^{(k-1)}_{N-4}-A^{(k)}_{N-2}},
\label{wynn_epsilon}
\end{eqnarray}
when $\alpha=1$.
Here the initial condition is given 
by $A^{(0)}_{N}=G(N)$ and $A^{(-1)}_{N}=\infty$. 
It was in ref.~\ref{Golinelli2} that 
this transformation was applied for the first time 
to estimate the $S=1$ Haldane gap. 
Note that when we take $\alpha=0$ in eq.~(\ref{wynn_epsilon}), 
the transformation is reduced to the the one called 
as the Aitken-Shanks process\cite{Shanks}. 
The Aitken-Shanks transformation was used 
in ref.~\ref{TSakai_MTakahashi}. 
Both of the transformations make us possible to accelerate 
the convergence of a finite sequence and to 
give a candidate for the extrapolated value.  
It is, however,  difficult to obtain a systematic error 
only within the framework of each transformation. 
Under such circumstances, the authors of ref.~\ref{Golinelli2} 
considered the variance of $\alpha$ 
within a successful acceleration of the convergence. 
We have found problems in the argument of estimating 
their systematic error in ref.~\ref{Golinelli2} 
when we examine the convergence of the $S=1$ finite-size gaps 
up to $N=24$ under the periodic boundary condition. 
We present the results of the table of the convergence for $\alpha=1$ 
in Table~\ref{gap_s1_pbc}.  
Note here that a part up to $N=22$ in Table~\ref{gap_s1_pbc} 
was reported and that the energy difference of $N=24$ and its 
posterity are new. 
Let us mention the problems while we are reviewing the procedure 
of ref.~\ref{Golinelli2}. 
\begin{table*}[tb]
\caption{Sequence of finite-size gaps for the $S=1$ case under the periodic boundary condition and the convergence acceleration based on Wynn's algorithm (eq.(\ref{wynn_epsilon}) with $\alpha=1$). }
\label{gap_s1_pbc}
\begin{tabular}{r|lc|cc|cc|cc|cc|c}
\hline
$N$ & $G(N)=A^{(0)}_{N}$ & $\xi^{(0)}_{N}$ 
& $A^{(1)}_{N}$ & $\xi^{(1)}_{N}$  
& $A^{(2)}_{N}$ & $\xi^{(2)}_{N}$  
& $A^{(3)}_{N}$ & $\xi^{(3)}_{N}$  
& $A^{(4)}_{N}$ & $\xi^{(4)}_{N}$ & $A^{(5)}_{N}$ 
\\
\hline
 2 & 2              &      &            &      &            &       & & & & & \\
 4 & 1              &      &            &      &            &       & & & & & \\
 6 & 0.720627362624 & 1.57 & 0.61232025 &      &            &       & & & & & \\
 8 & 0.593555254375 & 2.54 & 0.48753251 &      &            &       & & & & & \\
10 & 0.524807950414 & 3.26 & 0.44377567 & 1.91 & 0.43525877 &       & & & & & \\
12 & 0.484196469912 & 3.80 & 0.42557752 & 2.28 & 0.41798489 &       & & & & & \\
14 & 0.458965346938 & 4.20 & 0.41757427 & 2.43 & 0.41308940 & 1.59  & 0.41258393 & & & & \\
16 & 0.442795561359 & 4.50 & 0.41394088 & 2.53 & 0.41152416 & 1.75  & 0.41114647 & & & & \\
18 & 0.432221469865 & 4.71 & 0.41223978 & 2.64 & 0.41095437 & 1.98  & 0.41074380 & 1.57 & 0.41071215 & & \\
20 & 0.425210314459 & 4.87 & 0.41141374 & 2.77 & 0.41071416 & 2.32  & 0.41058977 & 2.08 & 0.41055478 & & \\
22 & 0.420515020390 & 4.99 & 0.41099554 & 2.94 & 0.41059985 & 2.69  & 0.41052336 & 2.38 & 0.41050133 &1.85 & 0.41049811 \\
24 & 0.417346883838 & 5.08 & 0.41077448 & 3.14 & 0.41054158 & 2.97  & 0.41049612 & 2.24 & 0.41048618 &1.59 & 0.41048426 \\
\hline
\end{tabular}
\end{table*}

Let us explain the procedure of ref.~\ref{Golinelli2} briefly.  
First, one considers the following decay lengths defined as
\begin{equation}
\xi^{(k)}_{N}=2/\log 
\left(
\frac{A^{(k)}_{N-4}-A^{(k)}_{N-2}}{A^{(k)}_{N-2}-A^{(k)}_{N}}
\right),
\label{xi}
\end{equation}
in order to examine converging behavior 
of the sequence $A^{(k)}_{N}$ for each step $k$.  
One should note that $\xi^{(0)}_{N}$ increases monotonically. 
This means that the convergence becomes slower 
when $N$ is getting larger. 
This is a source of difficulties 
in estimating the extrapolated gap.  
In order to overcome this difficulty, an acceleration is introduced. 
To examine whether the acceleration 
of $A^{(k)}_{N}$ for each $k$ from $A^{(k-1)}_{N}$ is 
successful or not, ref.~\ref{Golinelli2} investigates 
the following three conditions. 
\begin{description}
\item[I] $A^{(k)}_{N}$ for each $k$ is monotonic. 
\item[II] $\xi^{(k)}_{N}$ for each $k$ is 
monotonically increasing, namely, the following condition holds, 
\begin{equation}
\xi^{(k)}_{N+2} > \xi^{(k)}_{N}. 
\label{convergence_criteria2}
\end{equation}
\item[III] The following condition holds,
\begin{equation}
\xi^{(k+1)}_{N} < \xi^{(k)}_{N}. 
\label{convergence_criteria3}
\end{equation}
\end{description}
Conditions {\bf I} and {\bf II} suggest that 
properties of the initial sequence are preserved 
even though the acceleration is carried out. 
Condition {\bf III} means 
whether an element with a long decay length is successfully 
removed by the acceleration. 
ref.~\ref{Golinelli2} considered that $A^{(5)}_{N=22}=0.410498$ 
is successfully accelerated and 
that it is reliable as an approximate for the gap value. 
In ref.~\ref{Golinelli2}, 
the region of $\alpha$ where all the three conditions hold  
was found around $\alpha=1$ within data up to $N=22$ 
even when one applies the acceleration iteratively up to $k=5$. 
Finally, the obtained region of $\alpha$ gave a systematic error. 

Let us examine the situations in all the presently available data 
up to $N=24$. 
In Table~\ref{gap_s1_pbc}, $\xi^{(k)}_{N}$ up to $k=2$ 
are monotonically increasing.
However, one finds that $\xi^{(3)}_{N}$ does not show 
a monotonic $N$-dependence and that 
$\xi^{(4)}_{N}$ decreases with $N$. 
Condition (\ref{convergence_criteria2}) does not hold. 
This means that it is unclear whether 
the acceleration of $A^{(3)}_{N}$, $A^{(4)}_{N}$, 
and $A^{(5)}_{N}$ are successful or not. 
Next, we examine the behavior of $\xi^{(3)}_{N}$ and 
$\xi^{(4)}_{N}$ when we tune $\alpha$ around $\alpha=1$. 
The decreasing behavior of $\xi^{(3)}_{N}$ disappears 
around $\alpha=1.2$; 
however, $\xi^{(4)}_{N}$ is still decreasing. 
The present examination of data up to $N=24$ suggests  
that it becomes unclear 
whether $A^{(5)}_{N=22}=0.410498$ is appropriate 
as a reliable estimate or not 
according to the criteria of the above three conditions. 
At least for $\alpha=1$, the acceleration of $A^{(k)}_{N}$ 
up to $k=2$ seems to be successful. 
It is possible to use $A^{(2)}_{N}$ instead of $A^{(5)}_{N=22}$ 
as a reliable estimate. 
In this case, let us remember that 
$A^{(2)}_{N}$ is monotonically decreasing.  
This suggests that any of data in $A^{(2)}_{N}$ gives 
an upper bound for the true Haldane gap. 
Even though the value of $\alpha$ is tuned, 
one obtains only an assembly of upper bounds. 
There is no evidence to show that the true gap value 
is in the region of the assembly. 
Thus, tuning $\alpha$ is not an appropriate way to obtain 
a reliable error of the Haldane gap. 
Therefore, we have to develop another strategy 
without tuning $\alpha$ to achieve it. 

\begin{figure}[tb]
\begin{center}
\includegraphics[width=8cm]{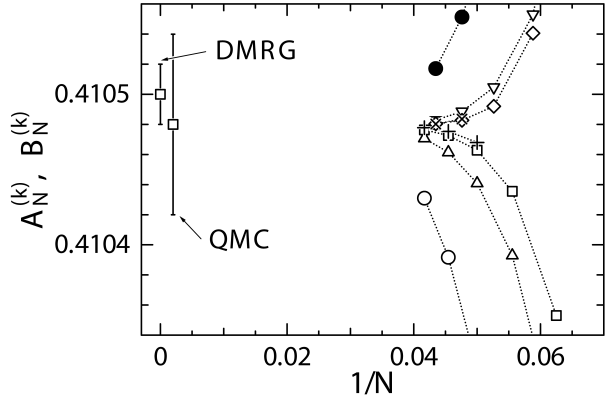}
\end{center}
\caption{The system size dependence of the sequences 
$A^{(k)}_{N}$ and $B^{(k)}_{N}$ 
under the twisted boundary condition. 
Open circles, triangles, squares, pluses denote 
$A^{(1)}_{N}$, $A^{(2)}_{N}$, $A^{(3)}_{N}$, and $A^{(4)}_{N}$, 
respectively. 
Closed circles, reversed triangles, diamonds, crosses denote 
$B^{(1)}_{N}$, $B^{(2)}_{N}$, $B^{(3)}_{N}$, and $B^{(4)}_{N}$, 
respectively. 
The estimates of the $S=1$ gap 
from the QMC method\cite{Todo_Kato_QMC} and 
from the DMRG one\cite{DMRG} are also presented  
near the left-hand ordinate. 
}
\label{a-b-comp}
\end{figure}

Next, we present the result of the convergence acceleration 
of our gap data under the twisted boundary condition. 
The table for $\alpha=1$ is shown in Table~\ref{gap_s1_tbc}. 
\begin{table*}[tb]
\caption{Sequence of finite-size gaps for the $S=1$ case under the twisted boundary condition and the convergence acceleration based on Wynn's algorithm (eq.(\ref{wynn_epsilon}) with $\alpha=1$). }
\label{gap_s1_tbc}
\begin{tabular}{r|lc|cc|cc|cc|cc|c}
\hline
$N$ & $G(N)=A^{(0)}_{N}$ & $\xi^{(0)}_{N}$ 
& $A^{(1)}_{N}$ & $\xi^{(1)}_{N}$  
& $A^{(2)}_{N}$ & $\xi^{(2)}_{N}$  
& $A^{(3)}_{N}$ & $\xi^{(3)}_{N}$  
& $A^{(4)}_{N}$ & $\xi^{(4)}_{N}$ & $A^{(5)}_{N}$ 
\\
\hline
 4 & 0.297769379131 &      &            &      &            &      &  &  &  &  & \\
 6 & 0.362613495315 &      &            &      &            &      &  &  &  &  & \\
 8 & 0.386237672978 & 1.98 & 0.39977728 &      &            &      &  &  &  &  & \\
10 & 0.396943190982 & 2.53 & 0.40581471 &      &            &      &  &  &  &  & \\
12 & 0.402443823536 & 3.00 & 0.40825701 & 2.21 & 0.40920572 &      &  &  &  &  & \\
14 & 0.405509288158 & 3.42 & 0.40936819 & 2.54 & 0.40998448 &      &  &  &  &  & \\
16 & 0.407315632794 & 3.78 & 0.40990703 & 2.76 & 0.41027702 & 2.04 & 0.41035307 &     &  &  & \\
18 & 0.408423139414 & 4.09 & 0.41017830 & 2.91 & 0.41039295 & 2.16 & 0.41043554 &     &  &  & \\
20 & 0.409122144092 & 4.35 & 0.41031824 & 3.02 & 0.41044086 & 2.26 & 0.41046285 & 1.81 & 0.41046801 &  & \\
22 & 0.409572951736 & 4.56 & 0.41039177 & 3.11 & 0.41046146 & 2.37 & 0.41047263 & 1.95 & 0.41047535 &  &  \\
24 & 0.409868488828 & 4.74 & 0.41043100 & 3.18 & 0.41047068 & 2.49 & 0.41047640 & 2.10 & 0.41047777 & 1.80 & 0.41047815 \\
\hline
\end{tabular}
\end{table*}
One can observe that all the above conditions hold. 
It is reasonable to consider that all of $A^{(k)}_{N}$ are 
successfully accelerated. 
The difference between the periodic and twisted 
boundary conditions is the direction of $A^{(k)}_{N}$; 
namely $A^{(k)}_{N}$ in Table~\ref{gap_s1_tbc} 
is monotonically increasing. 
This means that the data in $A^{(k)}_{N}$ are lower bounds. 
In order to obtain a reliable error only from the data 
under the twisted boundary condition, it is required 
to create another sequence that is monotonically decreasing. 
Here let us consider a new sequence defined as
\begin{equation}
B^{(k)}_{N+1}=
\frac{T_{N}S_{N+2}-T_{N+2}S_{N}}{S_{N+2}-S_{N}-T_{N+2}+T_{N}},
\label{quasi_reflective_seq}
\end{equation}
where $T_{N}=A^{(k)}_{N}$ and $S_{N}=A^{(k-1)}_{N}$. 
If $A^{(k)}_{N}$ is successfully accelerated 
from $A^{(k-1)}_{N}$, it is expected that 
the sequence $B^{(k)}_{N+1}$ is convergent 
from the side opposite to $T_{N}=A^{(k)}_{N}$ and 
$S_{N}=A^{(k-1)}_{N}$. (See appendix.)
In Table~\ref{upperbounds_s1}, we present 
the result of $B^{(k)}_{N}$ obtained from $A^{(k)}_{N}$ 
under the twisted boundary condition. 
One observes that all of $B^{(k)}_{N}$ are decreasing with increasing $N$. 
The difference of the dependences of $A^{(k)}_{N}$ and $B^{(k)}_{N}$ 
makes us know that the true gap value is between them. 
In this works, we employ $A^{(4)}_{N}$ and $B^{(4)}_{N}$ 
whose dependences are confirmed to be opposite 
to play safe. 
Namely, the $S=1$ Haldane gap is expected to be between 
$A^{(4)}_{N=24}=0.41047777$ and $B^{(4)}_{N=23}=0.41048023$. 
Therefore, our present consequence for the Haldane gap for $S=1$ is
\begin{equation}
0.4104789 \pm 0.0000013. 
\label{gap_s1_acceleration}
\end{equation}
This estimate agrees with the estimate 
from the QMC method\cite{Todo_Kato_QMC} and 
that from the DMRG one\cite{DMRG}. 
Note that our estimate (\ref{gap_s1_acceleration}) is 
more precise than any other estimates 
as long as the present authors know. 
Before finishing this paragraph, 
we illustrate $A^{(k)}_{N}$ and $B^{(k)}_{N}$ 
in Fig.~\ref{a-b-comp} so that one visually captures the features 
of these sequences, which are explained in the above. 

The sequence $B^{(k)}_{N}$ is a general way 
to estimate a reliable error; 
the case of the periodic boundary condition 
in Table~\ref{gap_s1_pbc} is applicable. 
In this case, $B^{(k)}_{N}$ is created 
from $S_{N}=A^{(1)}_{N}$ and $T_{N}=A^{(2)}_{N}$ 
because it is confirmed that 
$A^{(2)}_{N}$ is successfully accelerated from $A^{(1)}_{N}$. 
Note that in this case, $B^{(k)}_{N}$ is 
monotonically increasing and converging from the smaller side. 
Thus, $B^{(k)}_{N}$ gives a lower bound of the gap value.   
The data up to $N=24$ under the periodic boundary condition 
suggests that the Haldane gap is 
between $B^{(2)}_{N=23}$ and $A^{(2)}_{N=24}$; 
we have $0.41050 \pm 0.00005$ as an estimate 
in the periodic boundary condition. 
This result is also consistent with 
the estimate (\ref{gap_s1_acceleration}),
the one from the QMC method\cite{Todo_Kato_QMC}, and 
the one from the DMRG one\cite{DMRG}.  
This indicates that the present procedure makes it possible 
to estimate the gap value irrespective of boundary conditions.  
The difference of systematic errors between the cases 
of the periodic boundary condition and 
the twisted boundary condition originates 
from the characteristics of the initial sequences. 
In this meaning, the twisted boundary condition is better than 
the periodic boundary condition to estimate the Haldane gap.  
\begin{table}[tb]
\caption{Upper-bound sequence $B^{(k)}_{N}$ 
of the $S=1$ Haldane gap.}
\label{upperbounds_s1}
\begin{tabular}{r|c|c|c|c}
\hline
$N$ 
& $B^{(1)}_{N}$ 
& $B^{(2)}_{N}$ 
& $B^{(3)}_{N}$ 
& $B^{(4)}_{N}$ 
\\
\hline
9  & 0.41728860 &  &  &  \\  
11 & 0.41289925 &  &  &  \\  
13 & 0.41156232 & 0.41142826 &  &  \\  
15 & 0.41100869 & 0.41071649 &  &  \\  
17 & 0.41074765 & 0.41055312 & 0.41054058 &  \\  
19 & 0.41061764 & 0.41050468 & 0.41049203 &  \\  
21 & 0.41055134 & 0.41048859 & 0.41048271 & 0.41048358 \\  
23 & 0.41051712 & 0.41048286 & 0.41048036 & 0.41048023 \\  
\hline
\end{tabular}
\end{table}

\subsection{Cases for $S \ge 2$}

In this subsection, we estimate the Haldane gap for $S\ge 2$. 
We employ the twisted boundary condition 
to obtain the finite-size gap $G(N)$. 
Since the sequence $G(N)$ is increasing with $N$, 
it is easy to distinguish whether the gap survives or not 
in the limit of $N\rightarrow\infty$ 
even though the gap value is extremely small. 

Let us extrapolate our finite-size gaps for the $S=2$ case 
by eqs. (\ref{wynn_epsilon}) and (\ref{quasi_reflective_seq}). 
The result is summarized in Table~\ref{upperbound_S2}. 
One finds that data for small $N$ disturb the table. 
In Table~\ref{upperbound_S2}, $A^{(0)}_{4}$ and 
data originating from it are underlined. 
If we exclude these underlined data from the table, 
all other data do not disturb the table, which suggests 
that the convergence acceleration of the sequence $A^{(0)}_{N}$ 
starting from $N=6$ is successful. 
One can also find that $B^{(1)}_{N}$ without the underline 
show the dependence opposite to $A^{(k)}_{N}$. 
After excluding the underlined data, 
we do not have a sufficient number of data 
in $B^{(2)}_{N}$ to know its dependence; 
thus, we do not employ $B^{(2)}_{15}$ as an upper bound 
in this work. 
In addition, we cannot confirm whether $A^{(2)}_{14}$ and 
$A^{(2)}_{16}$ are successfully accelerated or not, 
because available $\xi^{(2)}_{N}$ originates from $A^{(0)}_{4}$.  
From the above reason, 
it is a careful and reliable judgment 
to consider that $\Delta(S=2)$ is between $A^{(1)}_{16}$ 
and 
$B^{(1)}_{15}$. 
Therefore our conclusion of the estimates for $\Delta(S=2)$ is 
\begin{equation}
0.0886 \pm 0.0018, 
\label{gap_s2_acceleration}
\end{equation}
for $S=2$. 
This estimate agrees with 
$\Delta = 0.08917 \pm 0.00004$ from the QMC simulation 
in ref. \ref{Todo_Kato_QMC} and 
$\Delta = 0.0876 \pm 0.0013$ from the DMRG calculation
in ref. \ref{Wang}.

\begin{table*}[tb]
\caption{Sequence of finite-size gaps for the $S=2$ case 
under the twisted boundary condition, 
the convergence acceleration based on Wynn's algorithm, 
and upper-bound sequence $B^{(k)}_{N}$. 
}
\label{upperbound_S2}
\begin{tabular}{r|cc|cc|c|cc|c}
\hline
$N$ 
& $10^{2} A^{(0)}_{N}$ & $\xi^{(0)}_{N}$ 
& $10^{2} A^{(1)}_{N}$ & $\xi^{(1)}_{N}$ 
& $10^{2} B^{(1)}_{N}$ 
& $10^{2} A^{(2)}_{N}$ & $\xi^{(2)}_{N}$ 
& $10^{2} B^{(2)}_{N}$ 
\\
\hline
 4  & \underline{3.60543243815}  & & & & & & &    \\  
 6  & 5.88636574630  & & & & & & &    \\  
 8  & 6.98099292140  & \underline{2.72} & \underline{7.9910262}   & & & & &    \\  
 9  & & & & & \underline{9.046111} & & &    \\  
10  & 7.57841496301  & 3.30 & 8.2962537   & & & & &    \\  
11 & & & & &            9.125346  & & &    \\  
12  & 7.93926067768  & 3.97 & 8.4896523  & \underline{4.38} & & \underline{8.6729401} & &    \\  
13 & & & & &            9.068141  & & & \underline{8.814100}   \\  
14  & 8.17386831221  & 4.65 & 8.6098758  & 4.21 & & 8.7252465 & &    \\  
15 & & & & &            9.021915  & & &            8.963622   \\  
16  & 8.33497991928  & 5.32 & 8.6881548  & 4.66 & & 8.7779957 & \underline{-237.32} &  \\  
\hline
\end{tabular}
\end{table*}

Next, we study the cases for $S \ge 3$. 
Our numerical results under the twisted boundary condition 
are presented in Table~\ref{upperbound_S345}. 
\begin{table*}[tb]
\caption{Finite-size gaps and decay lengths 
for the $S \ge 3$ cases under the twisted boundary condition. 
For $S = 3$, 
the convergence acceleration based on Wynn's algorithm 
is also applied. 
}
\label{upperbound_S345}
\begin{tabular}{r|cc|cc|cc|cc}
\hline
 & \multicolumn{4}{c|}{$S=3$} 
 & \multicolumn{2}{c|}{$S=4$} 
 & \multicolumn{2}{c}{$S=5$} 
\\ 
\hline
$N$ 
& $10^{3} A^{(0)}_{N}$ & $\xi^{(0)}_{N}$ 
& $10^{3} A^{(1)}_{N}$ & $\xi^{(1)}_{N}$ 
& $10^{4} A^{(0)}_{N}$ & $\xi^{(0)}_{N}$ 
& $10^{5} A^{(0)}_{N}$ & $\xi^{(0)}_{N}$ 
\\
\hline
 4  & 2.44233786473 &      &               &       & 1.29166071630 &      & 0.59777444079 &     \\  
 6  & 5.20317763910 &      &               &       & 3.52085939468 &      & 2.06995173926 &     \\  
 8  & 6.75739139783 & 3.48 & 8.75932105666 &       & 4.94293026662 & 4.45 & 3.12913003881 & 6.07  \\  
10  & 7.66279200087 & 3.70 & 8.92625320663 &       & 5.80259274045 & 3.97 & 3.79423190111 & 4.30  \\  
12  & 8.23027668731 & 4.28 & 9.18329148633 & -4.63 & 6.34716007738 & 4.38 & &     \\  
\hline
\end{tabular}
\end{table*}
One can observe that the decay length in $S=4$ and 5 
is not monotonically increasing. 
Although the decay length $\xi^{(0)}_{N}$ in $S=3$ is 
monotonically increasing, $\xi^{(1)}_{N}$ is negative. 
This indicates that 
$A^{(0)}_{4}$ and its posterity should be excluded 
from the table of acceleration 
according to the above criteria used in the case of $S=2$. 
After the exclusion, one cannot judge 
whether $A^{(1)}_{N}$ is successfully accelerated 
within the present data. 
Thus, we do not apply the above acceleration procedure 
to these finite-size gaps of $S \ge 3$ in this study. 

For $S \ge 3$, a serious behavior 
of the finite-size deviations appears 
when one draws a plot of $G(N)$ versus the inverse 
of the system size as a way that is usually applied. 
In this plot, the $1/N$-dependence of $G(N)$ reveals 
concave upwards for small sizes. 
For $S = 3$ and 4, the dependence becomes convex upwards 
for larger sizes. 
For $S = 5$, the dependence is still concave upwards 
in the range up to $N=10$. 
If the dependence is concave upwards, 
it is difficult for us to capture a converging behavior. 

Under these circumstances, 
we take a strategy composed of the following two steps. 
The first step is to draw a plot of $G(N)$ 
so that a shape which is concave upwards does not appear. 
The second step is to create a decreasing sequence 
from $G(N)$ that is an increasing sequence in the new plot.  

\begin{figure*}[tb]
\begin{center}
\includegraphics[width=16cm]{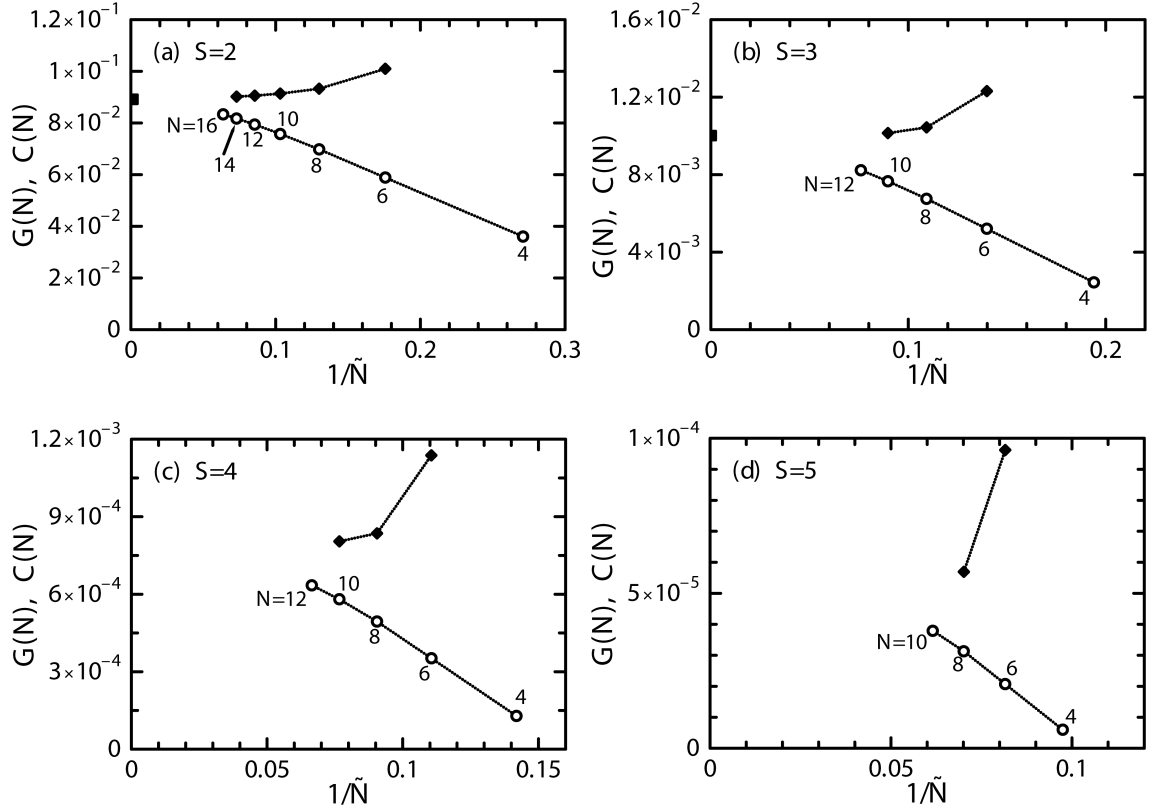}
\end{center}
\caption{Finite-size gaps $G(N)$ for $S=2$, 3, 4, and 5 
under the twisted boundary condition 
are shown by open circles. 
We determine a renormalized system size $\tilde{N}$ defined as 
$N+N_{0}$ so that three data for $N=4$, 6, and 8 
reveal a linear dependence; 
$N_{0}=-0.3091268$, 1.15226211, 3.04756114, and 6.25840221 
for $S=2$, 3, 4, and 5, respectively. 
For $S=2$ and 3, the results from QMC 
in ref.~(\ref{Todo_Kato_QMC}) are shown by closed squares. 
Diamonds represent upper bounds $C(N)$. 
Equation (\ref{fit_upper_bound_no_acceleration}) 
gives $C(N)$ from the finite-size gaps of system sizes 
$N$$-$2, $N$, and $N$$+$2. 
}
\label{s2-s5}
\end{figure*}

In order to carry out the first step, we here introduce 
a renormalized system size $\tilde{N}$ defined as $N+N_{0}$ 
so that three data for $N=4$, 6, and 8 reveal 
a linear dependence in the plot of $G(N)$ versus $1/\tilde{N}$. 
The results are depicted in Fig.~\ref{s2-s5}. 
One can observe the $1/\tilde{N}$-dependence of $G(N)$ 
is always convex upwards in every $S$ 
of Fig.~\ref{s2-s5}~(a)-(d). 
Note in Fig.~\ref{s2-s5}~(a) and (b) that 
$G(N)$ approaches an estimate obtained from QMC 
in ref.~(\ref{Todo_Kato_QMC}) for $S=2$ and 3. 
On the other hand, we do not have other estimates with which 
we can compare our $G(N)$ for $S=4$ and 5. 
All of $G(N)$ in Fig.~\ref{s2-s5} are increasing with $N$. 
Thus, our finite-size gaps $G(N)$ 
under the twisted boundary condition are appropriate 
to estimate the Haldane gaps even for $S \ge 2$. 
One finds that $G(N)$ gives lower bounds for the Haldane gaps. 

Next, we perform the second step. 
We focus on neighboring three data points of system sizes 
$N$$-$2, $N$, and $N$$+$2 in each panel of Fig.~\ref{s2-s5}. 
When we apply the fitting curve of 
\begin{equation}
y = C + D x^{E},
\label{fit_upper_bound_no_acceleration}
\end{equation}
to the three data points, 
it is possible to determine the parameters $C$, $D$, and $E$ 
uniquely for a given $N$. 
Then we use $C(N)$, $D(N)$, and $E(N)$ hereafter. 
Due to the above first step, $E(N=6)$ is necessarily the unity. 
Note that $D(N)$ is monotonically decreasing with increasing $N$ 
and that $E(N)$ is monotonically increasing. 
The result of $C(N)$ are also depicted 
at the corresponding $\tilde{N}$ in Fig.~\ref{s2-s5}.  
One can observe that $C(N)$ is monotonically decreasing 
with increasing $N$ in Fig.~\ref{s2-s5}~(a)-(d).  
At least for $S=2$ and 3, 
$C(N)$ seems to converge from the upper side 
to the gap value estimated 
from QMC calculations\cite{Todo_Kato_QMC}. 
It is reasonable to consider that 
the sequence $C(N)$ becomes an upper bound of the gap value. 
We choose $C(N)$ for the largest system sizes 
as the best upper bound in the analysis 
of Fig.~\ref{s2-s5}~(a)-(d). 
From the above argument of the lower and upper bounds 
of the gap value, we obtain 
\begin{eqnarray}
\Delta(S=2) &=& 0.0868 \pm 0.0034, 
\label{gap_s2_no_acceleration}
\\ 
\Delta(S=3) &=& 0.0092 \pm 0.0010, 
\label{gap_s3}
\\
\Delta(S=4) &=& 0.00072 \pm 0.00009, 
\label{gap_s4}
\\
\Delta(S=5) &=& 0.000047 \pm 0.000010. 
\label{gap_s5}
\end{eqnarray}
The estimate (\ref{gap_s2_no_acceleration}) for $S=2$ 
is consistent with the above result (\ref{gap_s2_acceleration}). 
We can also compare our estimate of $\Delta(S=3)$; 
our estimate agrees with 
$\Delta = 0.01002 \pm 0.00003$ from the QMC simulation 
in ref.~ \ref{Todo_Kato_QMC}.  
There is no other numerical estimate for $S \ge 4$ 
to the best of our knowledge. 
Our estimate (\ref{gap_s4}) and (\ref{gap_s5}) will be 
inspected in future if other methods become available.  

Note here that the analysis without convergence acceleration 
is also applicable to the case of $S=1$; 
the result is  
\begin{equation}
\Delta(S=1) = 0.41028 \pm 0.0042. 
\label{gap_s1_no_acceleration}
\end{equation}
This estimate is consistent 
with the gap (\ref{gap_s1_acceleration}). 
One of the differences is that 
the error of (\ref{gap_s1_no_acceleration}) is wider than 
the one of (\ref{gap_s1_acceleration}). 
The same situation appears 
between (\ref{gap_s2_no_acceleration}) and 
(\ref{gap_s2_acceleration}) in the case of $S=2$. 
These agreements for $S=1$ and 2 suggest that 
both the methods successfully lead to the common estimate 
that is irrespective of analyzing methods. 
Another difference is that the central value of the estimate 
(\ref{gap_s1_no_acceleration}) is slightly smaller 
than that of the estimate (\ref{gap_s1_acceleration}). 
One can also in the case of $S=2$ observe that 
the central value 
of the estimate (\ref{gap_s2_no_acceleration}) is smaller 
than that of the estimate (\ref{gap_s2_acceleration}). 
The estimate from the QMC simulation 
in ref.~ \ref{Todo_Kato_QMC} is larger than the central value 
of the estimate (\ref{gap_s3}) in the case of $S=3$. 
These facts suggest that the present method 
using eq. (\ref{fit_upper_bound_no_acceleration}) 
may give us the central value that is closer to the data 
of the original sequence. 
One of the reason may be that $C(N)$ 
of eq. (\ref{fit_upper_bound_no_acceleration}) 
is closer to the true quantity in the thermodynamic limit 
because $C(N)$ is a kind of extrapolated results. 
It is not so easy to know where within the error 
the true infinite-size quantity is 
only from the data of this work 
without referring other information. 
However, this question is beyond the present study 
because a primary purpose of this work is the development of 
a method that makes us possible to obtain 
a reliable systematic error within which there exists 
the true infinite-size quantity. 

\begin{figure}[tb]
\begin{center}
\includegraphics[width=8cm]{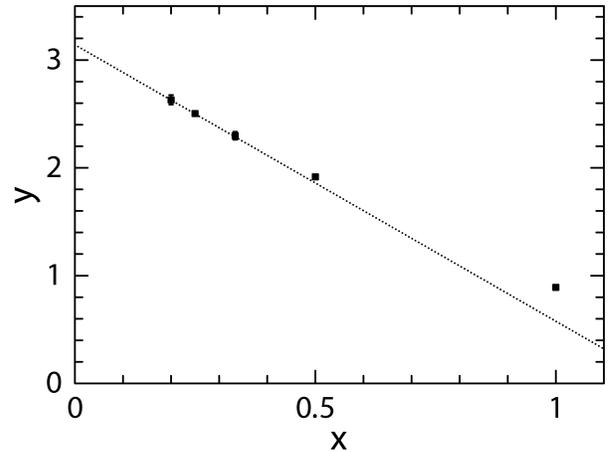}
\end{center}
\caption{Analysis of our estimates of Haldane gaps up to $S=5$ 
to confirm the asymptotic formula for large $S$. 
Errors for $S=1$ and 2 are smaller than the symbol size. 
The dotted line is the one with $\beta\sim 12.8$ 
obtained from the linear fitting of $S=4$ and 5. 
}
\label{asympt4j}
\end{figure}
Here, we examine the asymptotic formula of the Haldane gap, 
\begin{equation}
\Delta(S)= \beta |{\bf S}|^2 
\exp (-\pi |{\bf S}|), 
\label{asymptotic_formula}
\end{equation}
for $S\rightarrow\infty$ from our estimate of the gaps 
for finite $S$.  
In order to do it, 
we introduce new parameters $x = S^{-1}$ and 
$y = S^{-1} \log ( S^2/\Delta(S))$ 
when we take the amplitude of each spin to be $|{\bf S}|=S$ 
in the present analysis\cite{comment_amp_spin}. 
The asymptotic formula (\ref{asymptotic_formula}) 
is rewritten as 
\begin{equation}
y = \pi - x \log \beta. 
\label{analysis4asymptotic}
\end{equation}
Let us input our estimates of the Haldane gaps 
to $\Delta(S)$ in $y$ and plot the $x$ dependence of $y$. 
The result is depicted in Fig.~\ref{asympt4j}. 
One can find a linear behavior for finite but large $S$ 
up to $S=5$. 
We have fitted our data for $S=4$ and 5 with 
the straight line (\ref{analysis4asymptotic}); 
the best fit is produced by 
$\beta = 12.8 \pm 1.5$. 
The linear behavior suggests that 
the asymptotic formula (\ref{asymptotic_formula}) holds well 
for large $S$. 
Note here that if one uses the formula (\ref{asymptotic_formula}) 
and our estimate of $\beta$, 
one can predict the magnitude of the Haldane gap for $S>5$. 

\section{Summary and Remarks}

We have developed methods to estimate a systematic error 
in the extrapolation of finite-size data obtained 
from numerical diagonalizations of small clusters. 
The methods have been applied 
to study the Haldane gaps of 
the one-dimensional integer-$S$ Heisenberg antiferromagnet. 
We have demonstrated that the methods work well 
to the finite-size data under the twisted boundary condition. 
Our best estimate of the $S=1$ case is 
$\Delta=0.4104789 \pm 0.0000013$. 
We are also successful in obtaining the gaps up to $S=5$ 
even though the magnitudes of the gaps are extremely small. 
Our estimates of the gaps for large $S$ make us confirm 
that the asymptotic formula of the gap holds well. 
We have found that the twisted boundary condition 
is more appropriate 
than the periodic and open boundary conditions 
in order to extrapolate finite-size data 
of the Haldane gaps to the thermodynamic limit. 
However, the most appropriate boundary condition is not always 
the twisted boundary condition. 
The most appropriate boundary condition depends on the model and 
physical quantities\cite{different_case}. 
The present work strongly suggests that 
the examination of various boundary conditions is useful 
to know reliable quantities in the thermodynamic limit. 
It should be examined in future studies 
whether the methods work or not in extrapolations 
of other quantities, like the correlation functions of a system. 
The present method could give detailed information 
about the quantum state, which contributes much 
to a deeper understanding of properties of the system. 

Finally, it is noticeable that to produce a new sequence 
by eq.(\ref{quasi_reflective_seq}) is a quite versatile way. 
Only Wynn's transformation is employed in this work, 
but eq.(\ref{quasi_reflective_seq}) is available 
irrespective of acceleration methods. 
The parameter of sequences is not limited to the system size. 
Usefulness of this way for other parameters should be examined. 
In principle, the way of $B_{N}^{(k)}$ is applicable 
when the number of data in an original sequence $A_{N}^{(0)}$ 
is five at least because the direction of the dependence 
of $B_{N}^{(k)}$ can be checked. 
Note also that properties of the new sequence $B_{N}^{(k)}$ 
are related to whether the acceleration procedure is 
successful or not as we have mentioned in Appendix. 

\section*{Acknowledgment}
We wish to thank 
Prof.~K.~Hida, 
Dr.~K.~Okamoto, 
Prof.~T.~Sakai, 
Prof.~K.~Kusakabe, 
and Dr.~S.~Todo for fruitful discussions. 
This work was partly supported by Grants-in-Aid 
from the Ministry of Education, Culture, Sports, Science 
and Technology (No.~20340096). 
A part of the computations was performed using facilities 
of the Information Initiative Center, Hokkaido University 
and the Supercomputer Center, 
Institute for Solid State Physics, University of Tokyo. 

\appendix
\section{Convergence acceleration and new sequence}
Let us consider monotonic scalar sequences $T_{n}$ and $S_{n}$ 
that share the common limit $S$. 
\begin{table*}[t]
\caption{An application of the convergence acceleration 
to the sequence (\ref{sequence2pi}) having 
its limit as $\pi$. 
The decay lengths obtained from each sequence 
by eq.~(\ref{xi}) are also acompanied. }
\label{arcsin}
\begin{tabular}{r|lc|cc|cc|cc|cc|c}
\hline
$N$ & \mbox{}\hspace{7mm}$A^{(0)}_{N}$ & $\xi^{(0)}_{N}$ 
& $A^{(1)}_{N}$ & $\xi^{(1)}_{N}$ 
& $A^{(2)}_{N}$ & $\xi^{(2)}_{N}$ 
& $A^{(3)}_{N}$ & $\xi^{(3)}_{N}$ 
& $A^{(4)}_{N}$ & $\xi^{(4)}_{N}$ 
& $A^{(5)}_{N}$ 
\\
\hline
 2 & 2.922835737772 &      &            &      &            &      &            &      &  &  & \\
 4 & 3.032442077939 &      &            &      &            &      &            &      &  &  & \\
 6 & 3.081373479799 & 2.48 & 3.12083429 &      &            &      &            &      &  &  & \\
 8 & 3.106348882832 & 2.97 & 3.13238707 &      &            &      &            &      &  &  & \\
10 & 3.120142116779 & 3.37 & 3.13715611 & 2.26 & 3.13939334 &      &            &      &  &  & \\
12 & 3.128166065374 & 3.69 & 3.13932581 & 2.54 & 3.14068137 &      &            &      &  &  & \\
14 & 3.133009091490 & 3.96 & 3.14038271 & 2.78 & 3.14118662 & 2.14 & 3.14135408 &      &  &  & \\
16 & 3.136013726258 & 4.19 & 3.14092445 & 2.99 & 3.14140146 & 2.34 & 3.14149788 &      &  &  & \\
18 & 3.137917979434 & 4.39 & 3.14121336 & 3.18 & 3.14149861 & 2.52 & 3.14155251 & 2.07 & 3.14156654 &  & \\
20 & 3.139145542642 & 4.56 & 3.14137243 & 3.35 & 3.14154473 & 2.68 & 3.14157476 & 2.23 & 3.14158257 &  & \\
22 & 3.139947946854 & 4.70 & 3.14146233 & 3.50 & 3.14156750 & 2.83 & 3.14158434 & 2.37 & 3.14158853 & 2.02 & 3.14158978 \\
\hline
\end{tabular}
\end{table*}
According to ref. \ref{Brezinski}, 
the convergence of $T_{n}$ is faster than that of $S_{n}$ 
when the following condition is satisfied
\begin{equation}
\lim_{n\rightarrow\infty} \frac{T_{n}-S}{S_{n}-S} = 0. 
\label{def_acceleration}
\end{equation}
In a practical problem, however, 
lengths of the sequences are finite; 
$S$ is not a known quantity but the one that should be obtained. 
Therefore, the condition (\ref{def_acceleration}) cannot be 
examined directly. 
A substitute condition for (\ref{def_acceleration}) is given by 
\begin{equation}
\lim_{n\rightarrow\infty} 
\frac{T_{n}-T_{n+m}}{S_{n}-S_{n+m}} = 0. 
\label{def_acceleration2}
\end{equation}
for an arbitrary integer $m(> 0)$. 
eq. (\ref{def_acceleration2}) suggests that 
\begin{eqnarray}
1
&>&
\left|
\frac{T_{n}-T_{n+m}}{S_{n}-S_{n+m}}
\right|
\nonumber \\
&>& 
\left|
\frac{T_{n+\delta n}
     -T_{n+\delta n+m}}{S_{n+\delta n}
                       -S_{n+\delta n+m}} 
\right|
\nonumber \\
&>& 
\left|
\frac{T_{n+2\delta n}
     -T_{n+2\delta n+m}}{S_{n+2\delta n}
                        -S_{n+2\delta n+m}} 
\right|
>
\cdots,
\end{eqnarray}
for $n$ in the region where 
a converging behavior appears 
in $(T_{n}-T_{n+m})/(S_{n}-S_{n+m})$. 
Note here that 
this equation with $\delta n = m =2$ 
is related to eq. (\ref{convergence_criteria3}). 

Let us get back to eq. (\ref{def_acceleration}). 
Suppose that both $S_{n}$ and $T_{n}$ approach 
$S$ from the same side. 
The equation (\ref{def_acceleration}) suggests that 
\begin{equation}
\frac{T_{n+m}-S}{S_{n+m}-S} 
< 
\frac{T_{n}-S}{S_{n}-S},
\end{equation}
for sufficiently large $n$ and positive $m$. 
This inequality is rewritten as 
\begin{equation}
[(S_{n+m}-S_{n})-(T_{n+m}-T_{n})] S < 
T_{n} S_{n+m} - T_{n+m} S_{n}, 
\end{equation}
because $S_{n}$ is monotonic. 
When $T_{n}$ is obtained from $S_{n}$ 
through a successful acceleration procedure and  
when $S_{n}$ and $T_{n}$ are monotonically increasing, 
one can find 
\begin{equation}
S <  
\frac{T_{n} S_{n+m} - T_{n+m} S_{n}}{(S_{n+m}-S_{n})
-(T_{n+m}-T_{n}) }. 
\label{ineq_for_S}
\end{equation}
The right hand side of this inequality can be written to be 
\begin{equation}
S + \frac{T_{n}-S
-\frac{T_{n+m}-T_{n}}{S_{n+m}-S_{n}}(S_{n}-S)}{1
-\frac{T_{n+m}-T_{n}}{S_{n+m}-S_{n}}},
\label{limit_new_sequence}
\end{equation}
which is easily found to converge to $S$ 
in the limit of $n\rightarrow\infty$ 
with a help of the limit (\ref{def_acceleration2}). 

Let us consider the case of 
$T_{n}=A^{(k)}_{n}$, $S_{n}=A^{(k-1)}_{n}$, and $m=2$;  
The right hand side of the inequality (\ref{ineq_for_S}) 
becomes $B^{(k)}_{n+1}$. 
From the inequality (\ref{ineq_for_S}) and 
the limit of (\ref{limit_new_sequence}), 
$B^{(k)}_{n+1}$ defined as eq. (\ref{quasi_reflective_seq})
is found to converge to $S$ from the upper side.  
Note that the direction of the convergence of $B^{(k)}_{n+1}$ 
is opposite to that of $T_{n}$ and that of $S_{n}$. 

The above argument is applicable when 
$S_{n}$ and $T_{n}$ are monotonically decreasing. 
In this case, $B^{(k)}_{n+1}$ converges to $S$ 
from the lower side. 
The direction of the convergence of $B^{(k)}_{n+1}$ is 
opposite to that of $T_{n}$ and that of $S_{n}$ 
irrespective of the direction of $T_{n}$ and $S_{n}$. 

\begin{figure}[b]
\begin{center}
\includegraphics[width=8cm]{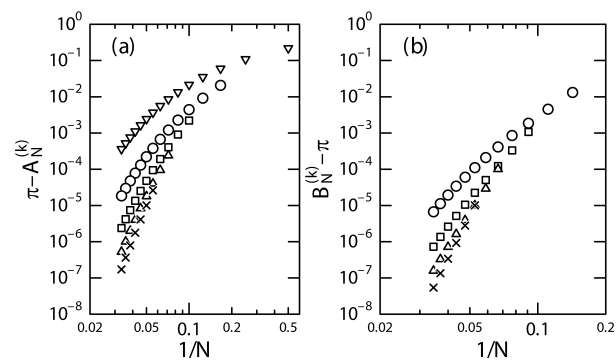}
\end{center}
\caption{
Log-log plots of (a) $\pi - A^{(k)}_N$ and (b) $B^{(k)}_N - \pi$ 
of sequence (\ref{sequence2pi}) for $N$ up to 30. 
Reversed triangles, circles, squares, triangles, and crosses 
in (a) 
denote $\pi-A^{(0)}_N$, $\pi-A^{(1)}_N$, 
$\pi-A^{(2)}_N$, $\pi-A^{(3)}_N$, and $\pi-A^{(4)}_N$, 
respectively. 
Circles, squares, triangles, and crosses in (b) 
denote $B^{(1)}_N-\pi$, $B^{(2)}_N-\pi$, $B^{(3)}_N-\pi$, 
and $B^{(4)}_N-\pi$, respectively. 
}
\label{pi-conv}
\end{figure}

Finally, a numerical example is presented 
for a demonstration of the procedure proposed in this paper. 
We consider the sequence $ \{ A^{(0)}_{N} \}
= \{ A^{(0)}_{2},A^{(0)}_{4},A^{(0)}_{6},A^{(0)}_{8},\cdots \} $ 
defined as
\begin{equation}
A^{(0)}_{N} = \sum_{k=0}^{N/2} \frac{(2k-1)!!}{(2k)!!}\frac{3}{2k+1}(\frac{\sqrt{3}}{2})^{2k+1}, 
\label{sequence2pi}
\end{equation}
where $(2k)!!=(2k)(2k-2)\cdots 4\cdot 2$ and 
$(2k-1)!!=(2k-1)(2k-3)\cdots 3\cdot 1$. 
This initial sequence is obvious to increase monotonically. 
It is easily understood that the limit of $N\rightarrow\infty$ 
is given by $3 \arcsin (\sqrt{3}/2)=\pi(=3.14159265\cdots)$ 
if one remembers the Taylor exapansion of $\arcsin x$ 
in a neighborhood of $x=0$. 
If we apply Wynn's transformation (\ref{wynn_epsilon}) 
and create a new sequence $B^{(k)}_{N+1}$ by means of 
eq.~(\ref{quasi_reflective_seq}) in the case of 
$T_{N}=A^{(k)}_{N}$ and $S_{N}=A^{(k-1)}_{N}$. 
The result is summarized in Table~\ref{arcsin}, 
Table~\ref{arcsin_b}, and Fig.~\ref{pi-conv}. 
One can confirm a successful acceleration of convergence 
of $A^{(k)}_{N}$ in Table~\ref{arcsin} 
from a judgement based on Conditions {\bf I}, {\bf II}, 
and {\bf III} in \S 3.1.2. 
A successful creation of $B^{(k)}_{N}$ 
which decreases monotonically with respect to $N$ 
can be observed in Table~\ref{arcsin_b}. 
Figure~\ref{pi-conv} is presented in order to confirm 
that the features of $A^{(k)}_{N}$ and $B^{(k)}_{N}$ hold 
for even larger $N$. 
One can find that each $B^{(k)}_{N}$ approaches 
the exact value from the upper side 
while $A^{(k)}_{N}$ approaches it from the opposite, 
namely lower, side. 
It is reasonable to consider that 
each datum of $A^{(k)}_{N}$ and $B^{(k)}_{N}$ 
is a lower bound and a upper bound, respectively,  
for the rigorous limit $\pi$. 
\begin{table}[b]
\caption{Upper-bound sequence $B^{(k)}_{N}$ 
calculated with eq.~(\ref{quasi_reflective_seq}) 
from $A^{(k)}_{N}$ in Table~\ref{arcsin}. }
\label{arcsin_b}
\begin{tabular}{r|c|c|c|c}
\hline
$N$ 
& $B^{(1)}_{N}$ 
& $B^{(2)}_{N}$ 
& $B^{(3)}_{N}$ 
& $B^{(4)}_{N}$ 
\\
\hline
 7 & 3.15479800 &  &  &  \\
 9 & 3.14614753 &  &  &  \\
11 & 3.14346184 & 3.14266173 &  &  \\
13 & 3.14244105 & 3.14192290 &  &  \\
15 & 3.14200462 & 3.14171496 & 3.14169304 &  \\
17 & 3.14180277 & 3.14164312 & 3.14162175 &  \\
19 & 3.14170394 & 3.14161507 & 3.14160274 & 3.14160271 \\
21 & 3.14165340 & 3.14160317 & 3.14159655 & 3.14159545 \\
\hline
\end{tabular}
\end{table}


\begin{thebibliography}{99}
\bibitem{history_Haldane_gap}
Various estimates of the Haldane gap of $S=1$ chain 
was extensively summarized in ref.~\ref{Golinelli2}. 
The case of $S=2$ was given in Table I of ref.~\ref{Wang}.
\bibitem{Householder} 
When system sizes are very small, the dimensions 
of the Hilbert space are also small; 
Lanczos calculations sometimes become unstable. 
In these cases, we perform the Householder diagonalizations 
instead.
\bibitem{different_case}
An example of a different model was reported 
in refs.~\ref{1D_Hubbard_chi} and \ref{2D_Hubbard_Drude}.
\end{thebibliography}
\end{document}